# Magnetization of superparamagnetics in the state of mechanical anisotropy

A. Ugulava[1], S. Chkhaidze[1], Sh. Kekutia[2], Z. Rostomashvili[3]

[1] I. Javakhishvili Tbilisi State University, 3, I.Chavchavadze av., 0179 Tbilisi, Georgia
[2] B.Chavchanidze Institute of Cybernetics, at Technical State University, 5, S.Euli str., 0186 Tbilisi, Georgia
[3] I. Gogebashvili Telavi State University, 1, University street, Telavi 2200, Georgia

**Abstract**

The internal energy of magnetic anisotropy for some nanoparticles dominates over the thermal energy even at room temperature. Strong magnetic anisotropy of nanoparticles can significantly affect the process of magnetization of the magnetic fluid. This influence is substantial if the system of nanoparticles is in a state of mechanical anisotropy in which the anisotropy axes of the particles have the same direction. In this work, it is shown that the magnetization curve of the magnetic fluid in a state of mechanical anisotropy is significantly different from that of Langevin. It is located between the Langevin and hyperbolic tangent curves and with increasing anisotropy takes progressively the hyperbolic tangent shape. It is also shown that in case of powder samples, the mechanical anisotropy leads to substantial quantitative changes in the Curie law.

## 1. Introduction

Over the past decades the field of research and development of magnetic nanomaterials has undergone dramatic changes. This is due to the development of efficient methods for preparation and stabilization of nanometer-scale magnetic particles (nanoparticles) as well as of physical methods for their investigation [1-4]. The increased interest of different-profile specialists in nanocrystals is aroused by various practical applications of nanomaterials. They are used in information record and storage systems, in new permanent magnets, in magnetic cooling systems, as magnetic sensors, etc.

When reducing particles to single-domain sizes (nanostructuring), and with preservation of spontaneous magnetization therein ($T < T_c$, $T_c$ is the Curie temperature), the influence of thermal fluctuations on the rotational dynamics of the magnetic moment *m* of a nanoparticle begins to grow. This type of random motion of the magnetic moment is called superparamagnetism [1-4], and the system consisting of a macroscopic number of magnetic nanoparticles - a superparamagnetic. One of the distinguishing features of superparamagnetics from conventional paramagnetic materials is that not individual atoms or molecules are carriers of the magnetic properties of elementary particles contained therein but magnetic nanoparticles containing a large number of atoms in the magnetically ordered state. The magnetic moments *m* of nanoparticles are much larger than magnetic moments of the single particles of a conventional paramagnetic of the

2order of only a few Bohr magnetons. Another distinguishing feature of superparamagnetics compared to conventional paramagnetics is related to the presence of the magnetic anisotropy energy of their particles.

If nanoparticles in a superparamagnetic are obtained by nanostructuring easy-axis ferromagnetics, they will also possess a magnetization easy axis (anisotropy axis). The anisotropy energy of uniaxial nanoparticles can be represented as [1-4]:

$$H_A(\theta) = A \sin^2 \theta, \qquad (1)$$

where $A$ is the anisotropy constant depending on the nanoparticle size, $\theta$ is the angle between the vector direction of the magnetic moment $\mathbf{m}$ of the nanoparticle and the anisotropy axis $\mathbf{n}$ (Fig.1). Usually, the anisotropy factor and the magnitude of the magnetic moment of the nanoparticle are represented as [1-4] $A = KV_m$ and $m = |\mathbf{m}| = M_S V_m$ where $K$ and $M_S$ are the densities of the corresponding quantities, $V_m = \dfrac{\pi}{6} d_m^3$ is the central magnetic sphere volume, and $d_m$ is the so-called "magnetic diameter" of the particle.

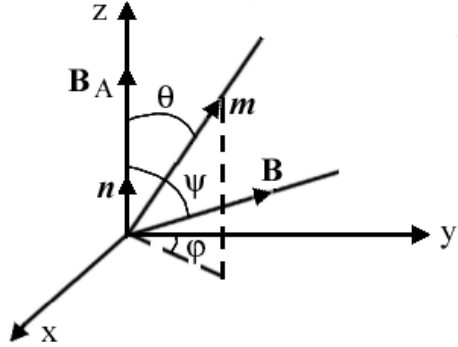

**Fig.1.** Orientation of the vectors of the magnetic moment $\mathbf{m}$ and magnetic induction $\mathbf{B}$ with respect to the easy-axis magnetization, $\mathbf{B}_A$ is the magnetic induction of the anisotropy field, $\psi$ is the angle between the easy-axis $\mathbf{n}$ and the magnetic induction vector $\mathbf{B}$, $\theta$ and $\varphi$ are the polar and azimuthal angles of the vector $\mathbf{m}$, respectively. At $\varphi \neq 0$ the angles $\theta$ and $\psi$ lie in different planes.

The magnetization process of a conventional "ideal gas" of paramagnetic particles possessing no internal magnetic anisotropy energy is well described by the classical Langevin theory. According to this theory, the macroscopic magnetization of the system of particles with the magnetic moment $\mu$ is given by the expression $M = N\mu L(B\mu/kT)$, where $L(x) = \operatorname{cth} x - 1/x$ is the Langevin function, $N$ is the number of particles per unit volume, $B$ is the magnetic induction of the external magnetic field, $k$ is the Boltzmann constant, $T$ is the absolute temperature. However, the presence of various forms of the anisotropy energy is characteristic for all magnetic nanoparticles. Its highest value ($K = 4.5 \cdot 10^5$ J/m$^3$) is observed in hexagonal-structure cobalt [5], and in magnetite ($Fe_3O_4$) nanoparticles it is lower by the order of magnitude ($K = 4.8 \cdot 10^4$ J/m$^3$) [3]. Nevertheless, by analogy with a conventional paramagnetic, the formula for the macroscopic magnetization of a supermagnetic consisting of magnetic nanoparticles is written in Langevin form by replacing $\mu$ by the nanoparticle magnetic moment $m$. Note that the resulting magnetization of a superparamagnetic does not depend on the coefficient $A$ which is by no means a small value for superparamegnetic nanoparticles. How to justify the validity of the application of the Langevin formula to the ensemble of magnetoanisotropic nanoparticles? To what changes in the magnetization curve may the polarization of anisotropy axes of nanoparticles lead? – These are the questions to be answered by this work.

## 2. General properties of magnetic nanoparticles

In practice, the ensemble of single-domain magnetic nanoparticles in the supermagnetic state is used as: a) a powder of magnetic nanoparticles [6-12], b) a system of magnetic nanoparticles in the solid-state matrix or in biological objects [13-16], and c) a liquid suspension of magnetic nanoparticles (magnetic fluid or ferrofluids [16-26]). In the first two cases, the axes of magnetic anisotropy of nanoparticles are randomly oriented and fixed ("frozen"). In the third case, the nanoparticles can rotate. Hereinafter, we will regard them as nanoparticles in the solid-state (cases a) and b)) and liquid matrices (case c)), respectively.

The Hamiltonian function of a uniaxial magnetic nanoparticle in the magnetic field has the form [1]

$$H(\theta,\psi,\varphi) = A\sin^2\theta - E(\cos\theta\cos\psi + \sin\theta\sin\psi\cos\varphi), \quad E = mB, \quad A > 0, \qquad (2)$$

The first term in formula (2) represents the single axis magnetic anisotropy energy (1), and the second term corresponds to the interaction of the magnetic moment $m$ of the particles with the magnetic field $B$.

In equilibrium the angles $\theta$ and $\psi$ determining the magnetic moment direction are taken from the energy minimum condition (2)

$$\frac{\partial H}{\partial \varphi} = E\sin\theta\,\sin\psi\,\sin\varphi = 0, \qquad (3)$$

$$\frac{\partial H}{\partial \theta} = A\sin 2\theta - E(-\sin\theta\,\cos\psi + \cos\theta\,\sin\psi\,\cos\varphi) = 0. \qquad (4)$$

In case of nanoparticles in a solid-state matrix, only the first of these two conditions can be fulfilled. Indeed, since the anisotropy axes of the nanoparticles $\psi$ in the solid-state matrix are randomly oriented and cannot rotate, and the angle $\theta$ varies in the range $0 < \theta \leq \pi$ in the magnetization process, neither of these angles is identically equal to zero, and the minimum condition (3) reduces to the identity equation $\varphi \equiv 0$, meaning that in equilibrium the angles $\theta$ and $\psi$ lie in the same plane. Condition (4) will take the form

$$A \sin 2\theta = E \sin(\psi - \theta). \tag{5}$$

The angles $\theta$ and $\psi$ are random variables. The randomness of the variable $\theta$ is due to thermal fluctuations, and the random nature of the variable $\psi$ is caused by the initial spread of the anisotropy axes of nanoparticles in the solid-state matrix. It is clear that there should not be any correlation between the two random variables. Therefore, minimum condition (5) for the nanoparticles in the solid-state matrix cannot be fulfilled. Then, applying the condition $\varphi = 0$ to the energy (2), we obtain for the Hamiltonian function of the magnetic nanoparticle in the solid-state matrix

$$H(\theta, \psi, \varphi = 0) = A \sin^2 \theta - E \cos(\theta - \psi). \tag{6}$$

As follows from the analysis [1] of the Hamiltonian function (6), the energy of a nanoparticle in the range $0 < \theta \leq \pi$ has two minima (or two potential wells) of different depth with the energy barrier between them of the order of $A$. For the magnetic moment to rotate (to transfer from one potential well to another), it is necessary to overcome this energy barrier. At temperatures below the blocking temperature ($T < T_B$) the magnetic moments are blocked in the potential wells and hence do not change their orientation. At $T > T_B$ over-barrier fluctuating transitions occur, and after some time $\tau_N$ the whole ensemble of nanoparticles forms a single magnetothermodynamic system. The formula for the characteristic time of the fluctuating over-barrier transitions at $a = A/kT \geq 1$ was first derived by Néel [1, 3]. The Néel relaxation process is related to the relaxation process inside the particle. Due to this relaxation the magnetic moment of the particle changes its direction, whereas the particle itself may remain fixed. Therefore, the Néel relaxation process is particularly important for magnetic nanoparticles in the solid-state matrix.

Generally speaking, at equilibrium, magnetic nanoparticles interact with each other. However, in particular conditions one can neglect this interaction and use the "ideal gas" model. Let us find out the conditions for this approximation to be possible.

Generally, only the central part of the nanoparticle has magnetic properties, and the particle itself is covered by a shell that has no magnetic properties. Therefore, taking a particle as a sphere, it is reasonable to introduce a total particle diameter $d$ along with the "magnetic" diameter $d_m$.

For magnetic particles a dipole-dipole interaction of the order of $\sim (\mu_0/4\pi)(m^2/d^3)$ is characteristic, where $\mu_0$ is the magnetic constant, and $m$ depends on the "magnetic" diameter. Then the condition for the magnetic interaction between the particles to be neglected can be written as

$$(\mu_0/4\pi)(m^2/d^3 kT) \ll 1. \tag{7}$$

To fulfill (7), it can be assumed that magnetic particles do not interact with each other and form an "ideal gas" of paramagnetic particles. The direct substitution shows that, for example, for spherical magnetite nanoparticles at room temperature the condition (8) is fulfilled if $d_m = 11$ nm and $d = 15$ nm ($V_m \approx 6.9 \cdot 10^{-25}$ m$^3$, $V \approx 18 \cdot 10^{-25}$ m$^3$, $m \approx 3.1 \cdot 10^{-19}$ A·m$^2$ $A = 3.3 \cdot 10^{-20}$ J). Below, we will use the "ideal gas" model of nanoparticles assuming that the system under study contains magnetite particles of such sizes. For a dimensionless barrier at room temperature we have $a = \dfrac{KV_m}{kT} \approx 8.2$.

### 3. Magnetization of nanoparticles with randomly directed anisotropy axes

Due to the random nature of the change in the angle variables $\theta$ and $\psi$ the Hamiltonian function (6) can be significantly simplified. Indeed, the second term in (6) depends on the difference of two independent random variables $\psi - \theta$. Obviously, this difference is also a random variable independent either of $\theta$ or of $\psi$. Let us introduce a new random variable $\xi$ by substitution $\psi - \theta \to \xi$. Then, the Hamiltonian (6) is represented in the form

$$H(\theta, \xi) = A \sin^2 \theta - E \cos \xi. \tag{8}$$

Following the general principles of statistical physics, perform standard calculations of the average magnetization [30]. Let us write a single-particle statistical integral for the Hamiltonian function (8)

$$z = (2\pi)^2 \int_0^\pi d\theta \sin\theta \int_0^\pi d\xi \sin\xi \, e^{-H(\theta,\xi)/kT}, \tag{9}$$

and then write a multiparticle statistical integral $Z = z^N/N!$, where $N$ is the number of particles in the sample unit volume. Calculating $Z$, a free energy expression can easily be obtained

$$F = -kT \ln Z, \tag{10}$$

and using it, the average macroscopic magnetization of a superparamagnetic

$$M = -\frac{\partial F}{\partial B}. \tag{11}$$

The corresponding statistical integral z is represented as a product of two integrals one of which depends only on the dimensionless anisotropy energy $a = A/kT$, and the other – only on the Boltzmann factor $b = Bm/kT$:

$$z = z_A z_B, \tag{12}$$

where

$$z_A = 2\pi \int_0^\pi e^{-a \sin^2 \theta} \sin \theta d\theta = 4\pi \frac{e^{-a}}{\sqrt{a}} Erfi(\sqrt{a}),$$

$$z_B = 2\pi \int_0^\pi e^{b \cos \psi} \sin \psi d\psi = 4\pi \frac{\sinh b}{b}. \qquad (13)$$

Here $Erfi(x) = \frac{\sqrt{\pi}}{2} \int_0^\pi \exp(y^2) dy$ is one of the forms of the probability integrals, $z_A$ is the statistical integral of the system of nanoparticles with internal magnetic anisotropy in the zero magnetic field [31], $z_B$ is the statistical integral of the Langevin paramagnetic "gas". Then, the multiparticle statistical integral of the two combined subsystems will have the form

$$Z = \frac{(z_A)^N (z_B)^N}{N!}, \qquad (14)$$

It is easily seen that the free energy of the system breaks up into two terms

$$F = -kT \ln Z = F_A + F_B, \qquad (15)$$

where

$$F_A = -kTN \ln z_A, \qquad F_B = -kTN \ln z_B. \qquad (16)$$

Then, with allowance for the statistical integrals (13) and (14), the average magnetization of the system (11) becomes dependent only on the second term $F_B$ of the free energy (15). After simple calculations we find:

$$M = -\frac{\partial F_B}{\partial B} = NmL\left(\frac{mB}{kT}\right). \qquad (17)$$

So, for the system of magnetic nanoparticles in the solid-state matrix we have obtained the known [3, 4] result (17) which shows that in spite of the anisotropic nature of the nanoparticle energy, the magnetization of the system does not depend on the anisotropy constant $A$ and is expressed through the Langevin function. This result is due to the combined action of two random processes - random distribution of the anisotropy axes of nanoparticles and temperature fluctuations of the magnetic moments.

### 4. Magnetization of nanoparticles with the polarized axes of anisotropy

For nanoparticles in a liquid matrix, along with Néel, there is another relaxation mechanism associated with the possibility of rotation of a particle (change in the variable $\psi$). This mechanism is characterized by the Brownian relaxation time (or the time of rotational diffusion).

$$\tau_B = \frac{3V\eta}{kT}, \qquad (18)$$

$\eta$ is the fluid viscosity. Along with the relaxation forces the particle in a magnetic field is also under the action of a magnetic force moment $\partial H(\theta,\psi)/\partial\psi = E\sin(\psi-\theta)$ that causes rotation of the anisotropy axis. Since the particle is able to rotate, the minimum condition (5) can be fulfilled. The analysis of the rotational motion of such a particle has shown [3] that under certain conditions the anisotropy axes can be parallel to the magnetic field induction ($\psi=0$). The resultant state is called the state of mechanical anisotropy. The characteristic times of rotation of the particle, or of the establishment of the mechanical anisotropy state have the form

$$\tau'_r = \frac{V}{V_m} \cdot \frac{6\eta}{M_S B} = \tau_r \frac{B_A}{B}, \quad \text{at } B \ll B_A, \tag{19}$$

$$\tau_r = \frac{V}{V_m} \cdot \frac{6\eta}{M_S B_A}, \quad \text{at } B \gg B_A, \tag{20}$$

where $B_A \equiv 2K/M_S = 2A/m$ is the magnetic induction of the anisotropy field.

For the mechanical anisotropy induced by the magnetic field to appear, the rotation time $\tau_r$ (or $\tau'_r$) should be lower than the rotational diffusion time $\tau_B$ ($\tau'_r \ll \tau_B$ ($B \ll B_A$) or $\tau_r \ll \tau_B$ ($B \gg B_A$). This last condition, with allowance for (19) and (20), takes the form [3]

$$B \gg B_A \gg B_r \tag{21}$$

where $B_r = \frac{kT}{M_S V_m}$.

Substituting $\psi=0$ into (6), we obtain for the Hamiltonian function in the state of mechanical anisotropy

$$H_f(\theta) = H(\theta,0,0) = A\sin^2\theta - E\cos\theta. \tag{22}$$

A single-particle statistical integral for this state has the form

$$z_D = 4\pi \int_0^\pi d\theta \sin\theta \exp\left[-H_f(\theta)/kT\right], \tag{23}$$

Integration of (23) gives

$$z_D = z_+ + z_-,$$

where

$$z_\pm = 4\pi \frac{\exp\left[-\left(a+\frac{b^2}{4a}\right)\right]}{\sqrt{a}} \left[\text{Erfi}\left(\sqrt{a} \pm \frac{b}{2\sqrt{a}}\right)\right]. \tag{24}$$

Writing a multi-particle statistical integral and performing transformations similar to those of Section 3, we get the expression for the magnetic polarization ($P_f = M/Nm$) in the presence of mechanical anisotropy

$$P_f(a,b) = -\frac{b}{2a} + \frac{1}{2\sqrt{a}} \frac{2 \cdot sh\, b}{e^b D(\sqrt{a} + b/2\sqrt{a}) + e^{-b} D(\sqrt{a} - b/2\sqrt{a})}, \qquad (25)$$

where $D(x) = e^{-x^2} \int_0^x e^{t^2} dt$ is the Dawson function [33].

Let us consider some limiting cases of this expression. Using the asymptotic expansion of the Dawson function $D(x \gg 1) \approx \frac{1}{2x} + \frac{1}{4x^3} + ...$, it is easy to obtain from (25) a polarization value corresponding to the saturated magnetization state $P_f(a, b \to \infty) = 1$ for the strong fields limit ($b \gg a$, $a \geq 1$). Using the same asymptotic expansion of the Dawson function, from (25) we obtain for the "deep wells" limit ($a \gg 1$)

$$P_f(a \gg 1, b) \equiv P_h(b) = \tanh b. \qquad (26)$$

This result has a simple physical interpretation: in the limit $a \gg 1$, nanoparticles are strongly "constrained" in two wells and there arises an analogy with the two-level system where the difference of the level populations at the state of thermal equilibrium is also determined by the hyperbolic tangent. However, for the magnetic fluid of the magnetite with considered parameters of the particles and at room temperature ($a \approx 8.2$) this limiting case is not fulfilled, and it is necessary to use the general formula (25).

Summing up, it can be concluded that the magnetization curve of the magnetic fluid of the magnetite in the region of strong magnetic fields $B \gg B_r$ has the shape $P_f(a,b)$, whereas the other part of this curve in weak fields $B \ll B_r$ has the Langevin shape $P_L(b) = L(b)$. In other words, the magnetization process begins according to the Langevin curve and reaches saturation following the curve $P_f(a,b)$. The transition from one curve to another is due to the fact that the magnetization process is accompanied by a gradual ordering of the anisotropy axes of nanoparticles. At the beginning of magnetization ($B \ll B_r$) the anisotropy axes of nanoparticles are random, and at the end ($B \gg B_r$) – are directed along the magnetic field. Thus, the magnetic fluid polarization can be represented as

$$P_F(a, B/B_r) = e^{-\frac{B}{B_r}} P_L(B/B_r) + \left(1 - e^{-\frac{B}{B_r}}\right) P_f(a, B/B_r). \qquad (27)$$

For the magnetite nanoparticles of the above sizes, at room temperature, we have $B_r \approx 0.01$ tesla. The magnetization curve $P_F$ is given in Fig.2. Since in the weak fields limit $P_F(a, B \ll B_r) \approx \frac{mB}{3kT}$, the mechanical polarization of the magnetic fluid does not lead to any significant changes in the Curie law.

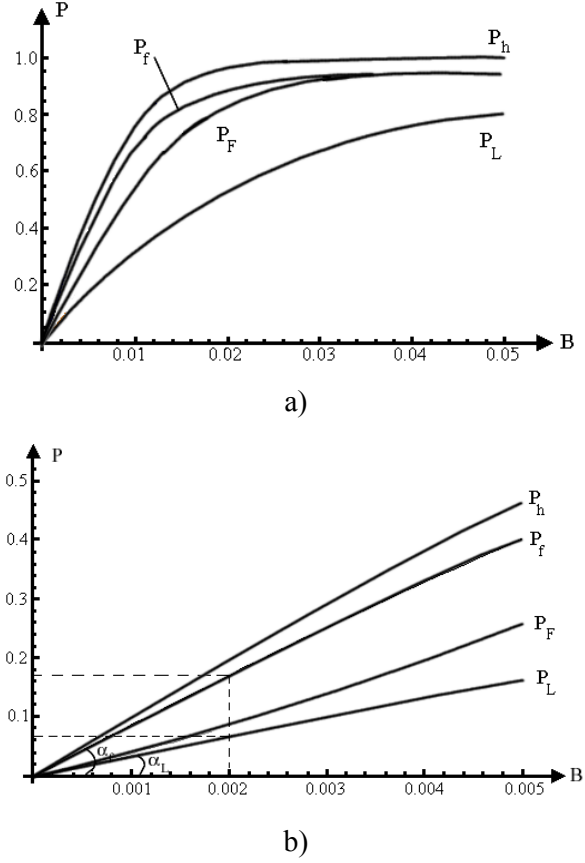

**Fig.2.** Curves of magnetic polarization of superparamagnetics.

a) Polarization curves: Langevin $P_L$, magnetic fluid $P_F$, powder in the state of mechanical polarization $P_f$, and hyperbolic tangent shapes $P_h$ plotted according to formulas (17), (25), (26) and (28) for magnetite nanoparticle sizes of $d_m$ = 11 nm and d = 15 nm and at room temperature ($a$ = 8.2, $b \approx 100B$).

b) Graphs are obtained from the curves of Fig.2 a) with scale magnification for low B. The slopes in these graphs show the Curie coefficient. For the case under consideration $\tan\alpha_f / \tan\alpha_L \approx 2.6$.

The state of mechanical anisotropy can also be created in powder superparamagnetics. Note that in such systems the chaotic direction of anisotropy axes is maintained by friction between the surfaces of the particles. Suppose we shake the container with the powder in a strong magnetic field. While shaking, the particles are released from the friction forces for a short time and acquire the ability to rotate freely. Assume that the conditions

$$\Delta t \gg \tau_B \gg \tau_r, \qquad \Delta t \gg \tau_B \gg \tau'_r. \qquad (28)$$

are fulfilled.

Then, for the time $\Delta t$, the state of mechanical anisotropy will be achieved in the system and will be maintained after this time. The rotation time of a nanoparticle can be evaluated from formula (20) supposing that $\eta$ is the air viscosity. Then, for the case under consideration $\tau_r \approx 2.1 \cdot 10^{-9}$ s. Based on this assessment, we can assume that condition (28) can be easily fulfilled. For the thereby obtained superparamagnetic the magnetization will occur following formula (25).

In the high-temperature approximation (weak fields, $b \ll 1$), in the first-order $b$ series expansion, we get from (25)

$$P_f(a, b \ll 1) = \left[ -\frac{1}{2a} + \frac{1}{2\sqrt{a}D(\sqrt{a})} \right] b. \qquad (29)$$

In terms of this expression the Curie law for a superparamagnetic in the solid-state matrix in the state of mechanical anisotropy can be represented as

$$M = C[1 + \Lambda(a)]B, \qquad (30)$$

where

$$C = \frac{m^2 N_0}{3kT} \qquad (31)$$

is the usual Curie constant, and

$$\Lambda(a) = \frac{\frac{3}{2\sqrt{a}} - \left( \frac{3}{2\sqrt{a}} + \sqrt{a} \right) D(\sqrt{a})}{\sqrt{a} D(\sqrt{a})} \qquad (32)$$

expresses its variation caused by mechanical anisotropy.

For powder samples of a magnetite of sizes $d_m = 11$ nm and $d = 15$ nm we have $D(\sqrt{8.2}) \approx 0.19$ and $\Lambda(8.2) \approx 1.64$. Usually, the number of nanoparticles is $N \approx 10^{15}$ $m^{-3}$ and therefore Curie coefficient $C \approx 0.64 \cdot 10^{-2}$ $A/m \cdot T$, and with allowance for the mechanical anisotropy it will be $M/B = C[1 + \Lambda(8.2)] \approx 1.7 \cdot 10^{-2}$ $A/m \cdot T$.

Thus, the formation of mechanical anisotropy in powder samples leads to an increase in the Curie coefficient by a factor of $r = 1 + \Lambda(8.2) \approx 2.6$.

Figure 2 shows that for the curves $P_f$, $P_h$ and $P_F$ the saturation occurs much earlier than for the Langevin curve $P_L = L(b)$. From comparison of the curves $P_L$ and $P_f$ it is clear that the shaking of the superparamagnetic powder in a strong magnetic field leads to a substantial change in the shape of the magnetization curve (2a), and consequently in the Curie coefficient (Fig.2b).

## Conclusion

The magnetization process of the macrosystem of superparamagnetics depends essentially on the orientation of anisotropy axes of nanoparticles. In powder samples the axes are oriented randomly and are "frozen". The consistent account of magnetic anisotropy showed that because of the random character of orientations of the axes of nanoparticles, the effect of the magnetic anisotropy is inhibited during magnetization. As a result, the magnetization curve has the Langevin shape.

In magnetic fluids, the system of nanoparticles is "unfrozen" and the axes of the particles can rotate under the influence of a magnetic field and random collisions of molecules of the fluid. Under the influence of a sufficiently strong magnetic field an alignment of the anisotropy axes occurs and a mechanical anisotropy state is established. The shape of the magnetization curve also changes considerably. At the beginning of the process the axes of nanoparticles are randomly oriented and the shape of the magnetization curve is similar to that of Langevin, and in the saturation area where the mechanical anisotropy state is achieved, it approaches the curve determined by formula (25). The formation of mechanical anisotropy in magnetic fluids does not lead to any changes in the Curie law.

Mechanical anisotropy can also be formed in powder superparamagnetics. Here, a substantial change in the Curie coefficient is observed.

## Acknowledgement

The authors acknowledge the funding from Shota Rustaveli National Science Foundation (Grant AR/96/3-250/13).

## References

[1] X. Batlle, A. Labarta. J. Phys. D. Appl. Phys. **35,** R15-R42 (2002).
[2] R. Skomski. J. Phys: Condens. Matter. **15,** R841-R896 (2003).
[3] M.I. Shliomis, Sov. Phys. Uspekhi **112,** 153-169 (1974).
[4] S. P. Gubin, Yu. A. Koksharov, G .B. Khomutov, G. Yu. Yurkov. Russian Chemical Reviews, **74 (6),** 489-520 (2005).
[5] S. P. Gubin and Yu. A. Koksharov. Inorganic Materials, **38,** 1085–1099 (2002).
[6] J. F. Löffler, J. P. Meier, B. Doudin, J. P. Ansermet, W. Wagner. Phys. Rev., B, **57,** 2915-2924 (1998).
[7] U. Gonser, C. J. Meechan, A. H. Muir, H. Wiedersich. J. Appl. Phys., **34,** 2373-2378 (1963).
[8] S. C. Abrahams, L. Guttman, J. S. Kasper. Phys. Rev., **127,** 2052-205 (1962).
[9] N. Saegusa, M. Kusunoki. Jpn. J. Appl. Phys., **29,** 876-877 (1990).
[10] T. Majima, T. Ishii, Y. Matsumoto, M. Takami. J. Am. chem.. Soc., **111,** 2417-2426 (1989).
[11] K. Haneda, Z. X. Zhou, A. H. Morrish, T. Majima, T. Miyahara. Phys. Rev., B, **46,** 13832-13837 (1992).
[12] S. Gangopadhyay, G. C. Hadjipanayis, B. Dale, c. M. Sorensen, k. J. Klabunde, V. Papaefthymiou, A. Kostikas. Phys. Rev., B, **45,** 9778-9887 (1992).

[13] V. Dupuis, J. Tuaillon, B. Prevel, A. Perez, P. Melinon, G. Guiraud, F. Parent, L. B. Steren, R. Morel, A. Barthelemy, A. Feri, S. Mangin, L. Thomas, W. Wernsdorfer, B. Barbara. J. Magn. Magn. Mater., **165**, 42-45 (1997).

[14] S. Honda, F. A. Modine, A. A. Meldrum, J. d. Budai, T. E. Haynes, L. A. Boatner. Appl. Phys. Lett., **77**, 711-713 (2000).

[15] S. Sun, S. Anders, H.F. Hamana, J. U. Thiele, J. E. E. Baglin, T. Thomson, E. E. Fullerton, C. B. Murray, B. D. Terris. J. Am. Chem. Soc., **124**, 2884-2885 (2002).

[16] R. B. Frankel, R. P. Blakemore, R. S. Wolfe. Science, **203**, 1355 (1979).

[17] S. Mornet, S.Vasseur, F. Grasset, E. Duguet. J. Mater Chem. **14**, 2116-2164 (2004).

[18] T. Upadhyay, R. V. Upadhyay, R. V. Mehta, V. K. Aswal, P. S. Goyal. Phys. Rev. B, **55**, 5585-5588 (1997).

[19] J. van Wonterghem., S. Morup, S. W. Charles, S. Wells. J. Colloid Interface Sci., **121**, 558-563 (1988).

[20] I. Volkov, M. Chukharkin, O. Snigerev, A. Volkov, S. Tanaka, C. Fourie. J. Nanopart.Res., **10**, 487-497 (2008).

[21] S. Laurent, D. Forge, M. Port, A. Roch, C. Robic, L. Vander Eist, R. N. Muller. Chem. Rev., **108**, 2064-2110 (2008).

[22] S. Laurent, S. Duts, U. O. Hafeli, M. Mahmoudi. Advances in Colloid and interface science, **166**, 8-23 (2011).

[23] Q. Song, Z. J. Zhang. J. Am. Chem. Sos. 126(19), 6164-6168, (2004).

[24] J. P. Chen, C. M. Sorensen, K. J. Klabunde, G. C. Hadjipanayis. Phys. Rev. B. **51. 17**, 11527-11532 (1995).

[25] J. P. Wilcoxon, E. L. Venturini, P. Provencio. Phys. Rev. B. **69**, 172402-172406 (2004).

[26] S. Taketomi, S. Chikazumi. Magnetic Fluids – principle and Application (Tokio, Nokkan Kogyo Shinbun, 1988), 272 p.

[27] W.T. Coffey, D. S. Crothers, J. L. Dorman, L. J. Geoghegan, Y. P. Kalmykev, J. T. Woldron, A. W. Wiokstead. Phys. Rev. B, **52,** 15951 (1995).

[28] W.T. Coffey, D. S. Crothers, J. L. Dorman, L. J. Geoghegan, Y. P. Kalmykev, J. T. Woldron, A. W. Wiokstead. J. Magn. Magn. Mater., **145**, L263 (1995).

[29] D. Leslie-Pelecky, R. D. Ricke. Chem. Mater. **8**, 1770 (1996).

[30] S. J. Blundell, R. M. Blundell. Consept in termal Physics. (Oxford, New York. 2010), 493 p.

[31] A. Ugulava, S. Chkhaidze, Sh. Kekutia, M. Verulashvili. Physica B, Cond. Mater., **454**, 249-252 (2014).

[32] W.F. Brown. J. Phys. Rev. **130**, 1677 (1963)

[33] M. Abramowitz, I.A. Stegun. Handbook of Mathematical Functions with Formules, Graphs and Mathematical Tables. New York Dover, 295-319, 1972.

A. Ugulava

Z. Rostomashvili

S. Chkhaidze

Sh. Kekutia